\documentclass[letterpaper,UKenglish]{llncs}
\usepackage{amsfonts}
\usepackage{amsmath}
\usepackage{hyperref}
\usepackage{microtype}

\usepackage{tikz}
\usetikzlibrary{automata,positioning, arrows}

\title{A Free Energy Foundation of Semantic Similarity in Automata and Languages}

\author{Cewei Cui and Zhe Dang}
\institute{School of Electrical  Engineering and Computer Science\\
Washington State University, Pullman, WA 99164,  USA\\
  \texttt{\{ccui,zdang\}@eecs.wsu.edu}}
\authorrunning{C.\, Cui and Z.\, Dang} 

\begin{document}

\maketitle

\begin{abstract}
 This paper develops a free energy theory from physics including the variational principles
for automata and languages and
 also provides algorithms to compute the energy as well as efficient algorithms for
estimating the nondeterminism in a nondeterministic finite automaton. This theory is then used as
a foundation to define a semantic
 similarity metric for automata and languages.   Since automata are a fundamental model
for all modern programs while languages are a fundamental model for the programs' behaviors,
we believe that the theory and the metric developed in this paper can be further used for real-word
programs as well.
 \end{abstract}

\section{Introduction}
Semantic similarity between two software systems plays  a central role in
studying software evolution, software plagiarism and in a more general context of semantic mining of an executable object. Clearly,  a syntactic metric
of such similarity defined on the source codes of two programs
(i.e., the source codes look similar) is far from being
 enough to catch the semantic
similarity. The reasons are not hard to see:
plagiarized software can have a completely different ``look"
even though its semantics is not much modified from the
original true copy. We shall also notice that the semantic metric that we are
looking for shall be an invariant on the dynamic behaviors (instead of the
source codes)  of the
software systems.

Automata are a fundamental model for all modern software systems while
languages (sets of words) are a model for the requirements
(i.e., behaviors specifications as sequences or words of events)
 of the systems. Hence, it is meaningful to study the semantic similarity
metric from the fundamental; e.g., finite automata and regular languages.
Such studies may provide a hint of inspiration for studying more
general software systems
and more importantly, finite automata themselves are useful as well in
software design (i.e., statecharts\cite{HAREL1987}).

We now take a new and physical
 view on a run of a finite-state program (i.e., a finite automaton $M$).  When $M$ runs, it receives input symbols while
each input symbols ``drives"  $M$ to transit from the current state to the next. We now imagine the automaton as a gas molecule while
each state that the run passes resembles an observation of the
molecule's physical position, speed, etc.  Hence, a run of $M$ corresponds to an observation sequence, called a microstate in physics, of the molecule. Clearly, the semantic  similarity metric that we look for must rely on an invariant
on the runs of an automaton $M$. This invariant can be found in the thermodynamic formalism
that provides a mathematical structure to rigorously deduce, in a many-particle system (e.g.\ gas),
from a microscopic behavior to a macroscopic
characteristic \cite{ruelle2004,sarig1999,walters1982,Gurevich1984}.
Even though a microstate is highly random,
the formalism resolves the challenge
 of how the highly random microstates in the system demonstrate almost stable
 macroscopic characteristic such as free energy.  Returning back to the automaton,  the free energy
computed can be treated as an invariant on the runs (i.e., dynamic behaviors) of the automaton $M$.
Notice that the free energy is measured on an equilibrium; being interpreted in a software system,
a program's long-term behavior is still highly dynamic and even random, however the dynamics itself is
invariant !  This thermodynamic view of software systems provides the physical
foundation of our semantic similarity metric.
However,  in order to build up the foundation, we need first develop a free energy theory for automata
and languages, and  the semantic similarity metric is then a direct and natural by-product of the theory. Notice that
the purpose of developing  the  theory is not limited to semantic similarity; it shall be considered
part of a new development in the traditional automata theory that is of more than 60 years of history.
We shall expect a wider range of applications of the theory itself, in addition to the similarity metric.

{\bf Related work}.  Delvenne \cite{Delvenne2011}  and Koslicki
and Thompson \cite{Koslicki14,Koslicki11} are among the first
 works to directly use the
thermodynamic formalism in Computer Science. In particular,  Delvenne \cite{Delvenne2011} computes page rank, using free energy on a random walk on
a graph,
 for web search while a ``missing link''
from a page to another is present. This work inspires us to study
 the
free energy of finite automata.
Koslicki
and Thompson \cite{Koslicki14,Koslicki11}  study topological pressure of a DNA sequence using a
thermodynamic formalism, where
an interested pattern on a DNA sequence is given a weight. This work inspires us to study free energy for formal languages
(sets of words).
Shannon's entropy has widely been considered to have a root in thermodynamics, e.g., Boltzmann equation.
In fact,  the notion of entropy rate  or information rate,   originally  by Shannon \cite{shabook}
 and later by Chomsky and Miller \cite{ChomskyM58}, that computes the
information quantity in a string (word), can be treated as a special case of free energy or topological pressure
in thermodynamic formalism,
 as pioneered in
Ruelle\cite{ruelle2004}, Walters \cite{walters1982}, Gurevich \cite{Gurevich1984},
Sarig \cite{sarig1999}, etc. Recently,  the classical formalism of Shannon's information rate has been used by
 the authors
in software analysis, testing,
and program security analysis \cite{CuiDFI16,LiD15,DangDFH15,IbarraCDF14}.  In particular,
 our previous paper \cite{CuiDF11}
 on the information rate  of a random walk of  a
graph has been recently used by Naval, Laxmi, Rajarajan, Gaur and Conti in malware detection \cite{Naval2015}.
 We are confident that  thermodynamic formalism can find its own applications in various areas
 of computer science,
 such as similarity in programs and
graphs \cite{cui2013similarity,Chartrand1998,dehmer2006similarity,elghawalby2008measuring,sokolsky2006simulation,zager2008graph}.
In particular, our previous work on program similarity \cite{cui2013similarity} is based on information rate and the Jaccard index.


The rest of the paper is organized as follows. We first briefly introduce
 an existing result in thermodynamic formalism, the variational principle \cite{ruelle2004}, in
its simplest form in physics. Then,
we develop a free energy theory for
automata and languages,
and, in various cases, provide algorithms to compute the quantity.
 Finally, we provide a semantic similarity metric based on the free energy theory.

\section{A free energy theory of automata and languages}

In Computer Science, a program's semantics can be interpreted as the set of all
 of
its behaviors.  In various  settings,  such a behavior can be simply a run
(such as a sequence of state transitions), or a sequence of I/O events.  Bearing thermodynamic formalism in
mind,  we understand the sequence as a microstate of the software system and ask a question:
\begin{quote}
Given the fact that there are so many (even infinite) microstates in a program,
and these microstates are highly ``probabilistic'' or random,
can we use thermodynamic formalism to help us understand some of
the program's macroscopic properties that are stable and {\em free of randomness}?
\end{quote}

In many ways,  some grand challenges in Computer Science have already partially
addresses the question. For instance,
it is well-known that unit testing
(roughly speaking, trying to understand part of an individual microstate)
 is nowadays relatively easy and can be done automatically in many cases \cite{Godefroid2005}. However,
system testing (roughly speaking, trying to understand a global or macroscopic property of the system under test)
is very difficult  \cite{Bertolino2007}.  In the research of P systems \cite{GPaun2002},  an evolution trace of a P system
can be considered as a microstate. However, a global property of such a highly dynamic system (which is Turing-complete)
 can be
very difficult to deduce and even be undecidable.  Model-checking \cite{Clarke2001,VardiW86} provides algorithms to verify
a global property of a finite state transition system through state exploration (or, roughly speaking, microstate exhaustive search), but
model-checking theories themselves also show how difficult verification can be because of, for instance, the
state-explosion problem \cite{Clarke2001} in a system of concurrent components.

The answer to the question above is yes but  it needs some work.
The set of behaviors is a language and we now
consider a language $L\subseteq \Sigma^*$ on a finite  and nonempty alphabet $\Sigma$ and a function
$\psi: \Sigma^\omega\to \mathbb{R}$.  In this paper, we consider a simplest form of $\psi$: for each
$x_0x_1x_2\cdots\in  \Sigma^\omega$,
$\psi(x_0x_1x_2\cdots)=U(x_0,x_1)$.  Herein, $U: \Sigma\times\Sigma\to\mathbb{R}$ is a given {\em cost function}
over words of length 2.  The intended purpose of  the cost function is to assign a cost to a pattern
in a word while the cost can be interpreted as, in a practical setting, an amount of
a resource, a priority level,  etc.

The cost is abstracted from ``potential''  in
 the
thermodynamic formalism, which we briefly explain
now.   An infinite word $\underbar x=x_0x_1x_2\cdots\in  \Sigma^\omega$ is a microstate in  thermodynamics. The microstate
evolves into $\sigma(\underbar x)=x_1x_2\cdots$,  $\sigma^2(\underbar x)=x_2\cdots,
\cdots ,  \sigma^n(\underbar x)= x_nx_{n+1}\cdots,$    etc.,  as discrete time $n$ evolves. Herein,  $\sigma$ is the shift-to-left operator defined in an obvious  way. The evolution from $\underbar x$ to $\sigma(\underbar x)$ is to break off
the first symbol $x_0$ in $\underbar x$ from the rest.  The break-off  needs energy
which is given by a pre-defined potential  $\psi(\underbar x)$. This explanation comes from
Sarig's lecture notes \cite{sarig-notes}
and we think it is a best way to illustrate the physics behind the mathematics.
  Hence, for an $n$-step evolution, the total energy needed
or the total potential possessed is
$(S^n\psi)(\underbar x)=_{\rm def}\sum_{0\le i\le n-1} \psi(\sigma^i(\underbar x)),$
 where $\sigma^0(\underbar x)=\underbar x$.

In  automata theory,
this can also be analogously understood as follows.
When
an infinite word
$\underbar x=x_0x_1x_2\cdots$ is read (symbol by symbol, from left to the right),  the reader  reads the first symbol
$x_0$, then
the second symbol $x_1$, etc.  Each such symbol-read consumes energy since essentially, it performs a shift-to-left operation
in the view of thermodynamics.

In thermodynamics,  a particle has high energy if it tends to evolve into one of  many choices of different microstates.
Similarly,  a  microstate is of high energy if it is the result of being chosen from many microstates.
This is characterized
 by the
Boltzmann equation $E=k_B\ln W$, where $W$ is the number of choices,
 $k_B$ is the Boltzmann
constant, and $E$ is the Boltzmann entropy. Hence,  $e^{(S^n\psi)(\underbar x)}$  now corresponds to the
number of choices for the $n$-step evolution of the microstate $\underbar x$ (while ignoring the Boltzmann constant).
Putting a summation over all
  $\underbar x$, we have the total number of ``choices'' for all microstates evolved for
$n$-steps,
$\sum_{\underbar x} e^{(S^n\psi)(\underbar x)}.$
But this formula, in general,  has no reason to give a finite value since there are simply
 an
uncountable number of
infinite sequences $\underbar x$. There are many ways to fix this.  One such way
 is due to Gurevich (see page 63 of \cite{sarig-notes}  and \cite{Gurevich1984})
where the summation is only over
the {\em periodic orbits} $\underbar x$ that satisfies $\sigma^n(\underbar x)=\underbar x$, considering the fact that
the $n$-step evolution really concerns the length $n$ prefix of $\underbar x$. As a result, the number of choices
 is now
$W_n=\sum_{\underbar x: \sigma^n(\underbar x)=\underbar x} e^{(S^n\psi)(\underbar x)}.$
Again, using
 the
Boltzmann equation, the total energy needed per step
for all microstates $\underbar x$ in $n$-step evolution
is ${1\over n}\ln W_n$.
Sending $n\to \infty$, we have $\lim {1\over n}W_n$ which is the
{\em (Gurevich) free energy}
\begin{equation}\label{eqq9003}
\lim_{n\to \infty} {1\over n} \ln \sum_{\underbar x: \sigma^n(\underbar x)=\underbar x} e^{(S^n\psi)(\underbar x)}.
\end{equation}
(The limit exists with a weak side condition (see Proposition 3.2 in \cite{sarig-notes}). For the mathematics,
see page 63 of \cite{sarig-notes}  and \cite{Gurevich1984}).

Coming back to Computer Science, we modify the formula in (\ref{eqq9003}) so that the summation is over the (finite)
words $w$ of length $n$ in a language
$L$ instead of $\omega$-words $\underbar x$. To do this, we first
consider $w=x_0\cdots x_{n-1}$ which is the prefix (with length $n$) of $\underbar x=x_0\cdots x_{n-1}\cdots$.
For the term $(S^n\psi)(\underbar x)$ in (\ref{eqq9003}),
we modify
it slightly into
\begin{equation}\label{eqq8001}
 (U)(w) =_{\rm def}\sum_{0\le i< n-1} U(x_i,x_{i+1})
\end{equation}
 while, as we have mentioned earlier, the $\psi$ now takes the special form
$\psi(x_0x_1\cdots)=U(x_0,x_1)$. Then
the periodic orbits in (\ref{eqq9003}) can be safely replaced with words of
length $n$.
 We thereby obtain
the definition of  {\em (Gurevich) free energy of language}  $L$:
\begin{equation}\label{eqq9004}
 {\rm G}_U(L)=\limsup_{n\to \infty} {1\over n} \ln \sum_{w\in L, |w|=n} e^{(U)(w)}.
\end{equation}
(By convention, $\ln 0=0$. The limit superior can be replaced by limit  in many cases, e.g., when $L$ is prefix-closed and regular, as we show later.)

Intuitively, the free energy  ${\rm G}_U(L)$ characterizes the average ``cost'' per symbol of words in
$L$ with respect to the cost function $U$.  A particularly interesting example is to interpret the cost as ``uncertainty'', as shown in a later example.
 A special case is when $U=0$.  In this case, the free energy
is simply (modulo a constant)
 the information rate of $L$ that was originally proposed by Shannon \cite{shabook} and Chomsky and Miller \cite{ChomskyM58}, and
more recently studied in \cite{CuiDFI14}.
We shall notice that, once $U$ is given, the free energy of $L$ is a constant and its definition does not involve
any probabilistic arguments.

 We explore
how to compute the free energy defined in
(\ref{eqq9004})   for  various classes of languages. We shall first point out that the  free energy is not computable
in general, even for simple classes of languages.
 \begin{theorem}\label{diff}
The  free energy  for language $L'=(\Sigma^*-L)\Sigma^*$, where $L$ is a context-free language, is not computable.
\end{theorem}

Later, we will show that the  free energy  is computable for regular languages.  The proof needs the variational principle in thermodynamics, which is briefly introduced as follows.

Consider the compact metric space whose topology is
 generated from the cylinder sets in the form of $[x_0\cdots x_{n-1}]=\{\underbar x: \underbar x=x_0\cdots x_{n-1}\cdots\}$, and uses  metric
$d(\underbar x, \underbar y)= 2^{-min\{i: x_i\ne y_i\}}$, where
$\underbar x=x_0\cdots x_{n-1}\cdots$ and $\underbar y=y_0\cdots y_{n-1}\cdots$.
Let $\mu$ be a probability measure over the space
 that is invariant
 under
the Markov shift $\sigma$.
In plain English, $\mu$ defines a Markov chain over the finite state space $\Sigma$, noticing that, herein, a symbol
in $\Sigma$ is a state
 of
the Markov chain.
For the given $\mu$, one can define the {\em free energy} with respect to
 $\mu$ as the quantity
$H_\mu+\int \psi d\mu,$
where $H_\mu$ is the Kolmogorov-Sinai entropy of the Markov chain $\mu$
(intuitively, it quantifies the average randomness, called entropy,
 on one step of the Markov chain), and
$\int \psi d\mu$ is the average potential on one step of the Markov chain.

 In a thermodynamic system,
nature tends
to make particles move in a way that maximizes the free energy \cite{sarig-notes}  as
\begin{equation}\label{eqq9008}
\sup_\mu~\{H_\mu+\int \psi d\mu\},
\end{equation}
which is called the {\em free energy} (or called Gurevich pressure)
 of the aforementioned
Markov shift $\sigma$ with potential $\psi$. We shall point out that the Markov shift itself does
not contain any probabilities.  For simplicity,
we omit much of the mathematics behind the definition which can be found in \cite{sarig-notes}.

One of the most important achievements in thermodynamic formalism establishes the variational principle \cite{ruelle2004}.
When interpreted on periodic orbits, it
says that the free energy, defined in (\ref{eqq9003}), is indeed the
 free energy on the Markov shift (again, with a side condition--see Sarig's notes\cite{sarig-notes}-- which we omit here for simplicity.):
\begin{equation}\label{eqq9009}
\lim_{n\to \infty} {1\over n} \ln \sum_{\underbar x: \sigma^n(\underbar x)=\underbar x} e^{(S^n\psi)(\underbar x)}=\sup_\mu~\{H_\mu+\int \psi d\mu\}
\end{equation}
The supremum
on the RHS of (\ref{eqq9009}) is achieved by an equilibrium probability measure $\mu^*$
(called Parry measure), which
can be computed from a nonnegative matrix, called Gurevich Matrix  \cite{Gurevich1984}, constructed from the definition of $\psi$ when $\psi$ is defined as $U$, mentioned earlier. In particular,  the
LHS of (\ref{eqq9009})  can be computed from the Perron-Frobenius eigenvalue of the matrix \cite{Gurevich1984}.

We now generalize the free energy  from Markov shift to a finite automaton. Let $M$ be a nondeterministic finite automaton (NFA)  with finitely many states specified by  $Q$ and with alphabet $\Sigma$.  Transitions in $M$ are specified by
a set $T\subseteq Q\times \Sigma\times Q$, where each transition $t\in T$ in the form of $(p,a,q)$, or simply written
$p\stackrel{a}{\to}q$,
means that $M$ moves from state $p$ to state $q$ on reading input symbol $a$.
A run is a sequence of transitions
\begin{equation}\label{eqq9010}
\tau=(p_0,a_0,p_1)(p_1,a_1,p_2)\cdots (p_{n-1},a_{n-1},p_n),
\end{equation}
for some $n$, satisfying
$p_0\stackrel{a_0}{\to}p_1\stackrel{a_1}{\to}\cdots  \stackrel{a_{n-1}}{\to}p_n$.
In $M$, there is a designated initial state $q_{\rm init}$ and a number of designated accepting states $q_{\rm accept}\in F
\subseteq Q$. The run $\tau$ is initialized if $p_0$ in (\ref{eqq9010}) is the initial state. It is an accepting run if
it is initialized and the last state $p_n$ in (\ref{eqq9010}) is an accepting state. In this case, we say that the word
$a_0\cdots a_{n-1}$ is accepted by $M$.  As usual, we use $L(M)$ to denote the language accepted by $M$.
$M$ is deterministic (i.e.\ a DFA) if, for each $p$ and $a$, there is at most one $q$ such that
$p\stackrel{a}{\to}q$.
It is well-known that an NFA can be converted into a DFA such that both automata accept the same language.

Throughout the paper, we assume that $M$ is cleaned up. That is, all the states are dropped from $M$ whenever it
 cannot
be reached from the initial state or it
 cannot
reach an accepting state.

We now associate a cost function $V: T\to \mathbb{R}$ which assigns a cost value to every transition in $M$.
 We
write $M_V$ for the $M$ associated with
 cost function $V$, called an NFA with cost.

We first assume that $M$ is strongly connected. That is,  every state can reach every state in $M$. More precisely,
for each $p$ and $q$, there is a run in the form of (\ref{eqq9010}) with $p_0=p$ and $p_n=q$.
We now define the free energy of $M_V$.
We
 note
that the results in \cite{Gurevich1984} are defined on strongly connected
graphs
 only.
However, a finite automaton $M$ is not, strictly speaking,
 a graph since there could be multiple
transitions from
 one
state to another (while in a graph there is at most one edge from a node to another).
Therefore, we first need to carefully translate the automaton $M$ into a Markov shift (a graph) as follows.
Let $\Theta$  be the set of all infinite sequences
in the form of
\begin{equation}\label{eqq9011}
\underbar t=p_0(p_0,a_0,p_1)p_1(p_1,a_1,p_2)p_2\cdots p_{n-1} (p_{n-1},a_{n-1},p_n)p_n\cdots
\end{equation}
or
\begin{equation}\label{eqq4011}
\underbar t=(p_0,a_0,p_1)p_1(p_1,a_1,p_2)p_2\cdots p_{n-1} (p_{n-1},a_{n-1},p_n)p_n\cdots
\end{equation}
satisfying $p_0\stackrel{a_0}{\to}p_1\stackrel{a_1}{\to}\cdots  \stackrel{a_{n-1}}{\to}p_n\cdots$
(we shall note that $\underbar t$ may not start from the initial state of $M$).
In terms of
 the
thermodynamics formalism, we define a potential function $\psi$ such that for each
$\underbar t$ in $\Theta$,
  its potential $\psi(\underbar t)=V(p_0,a_0,p_1)$ if  $\underbar t$ is in the form of (\ref{eqq9011});
$\psi(\underbar t)=0$ if $\underbar t$ is in the form of (\ref{eqq4011}).
We can similarly define a compact metric space over
$\Theta$ whose topology is generated by cylinders  and the metric $d$ defined earlier. Let $\mu$ be a
$\sigma$-invariant probability
 measure.  Now
the Markov shift $\sigma$ on $\Theta$ defines
a graph $\hat M$ as follows. For all $p, a,$ and $q$,
$\hat M$ has node $p$, node $q$, node $(p,a,q)$,  and edges from
node $p$ to node $(p,a,q)$ and from
node $(p,a,q)$ to node $q$,   iff 
 $(p,a,b)$ is a transition in $M$. Clearly, $\hat M$ is a strongly connected graph.

The free energy ${\cal E}(\hat M)$
 is defined as the quantity in
(\ref{eqq9008}). It is known that \cite{Gurevich1984} the free energy can be computed as follows.
Suppose that the graph $\hat M$ has $k$ nodes indexed with $1,\cdots, k$.
For a node $q$ (resp. node $(p,a,q)$) in $\hat M$, we use $[q]$
(resp. $[(p,a,q)]$) for its index.
We now construct a $k\times k$ matrix ${\bf M}$, called the Gurevich matrix,
as follows.
For each $i$ and $j$,
\begin{itemize}
  \item ${\bf M}_{ij}=0$
		if there is no edge from node $i$ to node $j$ in $\hat M$;
  \item ${\bf M}_{ij}=e^{V(p,a,q)}$
		if there is an edge from node $i$ to node $j$ in $\hat M$ and $i=[p]$ and $j=[(p,a,q)]$ for some $p,a,q$;
  \item ${\bf M}_{ij}=e^0$
		if there is an edge from node $i$ to node $j$ in $\hat M$ and $i=[(p,a,q)]$ and $j=[q]$ for some $p,a,q$.
\end{itemize}
Clearly, ${\bf M}$ is a non-negative and irreducible (since $M$ is strongly connected) matrix.
 Let $\lambda$ denote the spectral radius of ${\bf M}$, which is obtained as the largest positive real eigenvalue of ${\bf M}$,
 which is called
the Perron-Frobenius eigenvalue of ${\bf M}$.
Finally, according to
\cite{Gurevich1984}, the free energy ${\cal E}(\hat M)$  can be
efficiently computed as $\ln\lambda$.
\begin{lemma}\label{connected} (Gurevich theorem)
${\cal E}(\hat M)=\ln\lambda$, where $\lambda$ is the Perron-Frobenius eigenvalue of the
Gurevich matrix ${\bf M}$.
\end{lemma}
Finally, {\em the free energy ${\cal E}(M_V)$ for finite automaton} $M$ with
 cost function $V$ is defined as $2\cdot {\cal E}(\hat M)$. \footnote{The factor $2$, intuitively,
 comes from the fact that we ``stretch'', by a factor of 2,
 a run in finite automaton $M$ to correspond it to a walk in
graph $\hat M$. A
 somewhat
more efficient way to compute
 ${\cal E}(M_V)$ is to
construct
 an
$m\times m$ Gurevich matrix ${\bf M}'$
where $m$ is the number of states in $M$ such that
 ${\bf M}'_{ij}=0$ if there is no transition from $p_i$ to $p_j$ in $M$,
else
${\bf M}'_{ij}=\sum_{a:(p_i,a,p_j)\in T} e^{V(p_i,a,p_j)}$. Herein, $p_1,\cdots, p_m$ are all
 states in $M$.
One can show
 that
${\cal E}(M_V)=\ln\lambda'$ where $\lambda'$ is the
Perron-Frobenius eigenvalue of ${\bf M}'$. We omit the details.}

We shall
 note
that the definition of ${\cal E}(M_V)$  does not mention the initial and accepting states
of $M$. We now bring in those states and prove the variational principle for finite automata.
To do this, we first use a finite sequence to represent a periodic orbit, as we did in
(\ref{eqq8001}).  That is, for a run $\tau$ in (\ref{eqq9010}) with length $|\tau|=n$,
we define
\begin{equation}\label{eqq5001}
 (V)(\tau) =_{\rm def}\sum_{0\le i< n} V(p_i, a_i, p_{i+1}).
\end{equation}
In particular, we use $R$ to denote the set of all runs of $M$
and $A$ to denote the set of all accepting runs of $M$; herein,
both sets are languages on transitions in $M$.
 \begin{theorem}\label{NFAVP} (Variational Principle for Strongly Connected NFA) Let $M$ be an NFA that is strongly connected and with cost function $V$
on transitions. Then 
the following equations hold:
  \begin{equation}\label{eqq8977}
{\rm G}_V(R)=\lim_{n\to \infty} {1\over n} \ln \sum_{\tau\in R, |\tau|=n}
  e^{(V)(\tau)}={\cal E}(M_V).
\end{equation}
and
\begin{equation}\label{eqq8976}
{\rm G}_V(A)=\limsup_{n\to \infty} {1\over n} \ln \sum_{
\tau\in A, |\tau|=n}
 e^{(V)(\tau)}={\cal E}(M_V).
\end{equation}
In particular when every state in $M$ is an accepting state (hence $L(M)$ is prefix closed),
\begin{equation}\label{eqq8975}
{\rm G}_V(A)=\lim_{n\to \infty} {1\over n} \ln \sum_{
\tau\in A, |\tau|=n}
 e^{(V)(\tau)}={\cal E}(M_V).
\end{equation}
 \end{theorem}

 In general, $M$ is not necessarily strongly connected. However, it is well-known
(using the linear-time Tarjan algorithm) that
$M$ can be uniquely partitioned into a number of components $M^1,\cdots,M^k$, for some $k\ge 1$,
such that
\begin{itemize}
\item each $M^i$ is strongly connected, or it is
 singleton (i.e.\ it contains only one state that does not have a self-loop transition), and
\item each $M^i$ is maximal (i.e.\ the above condition is no longer true if it is enlarged).
\end{itemize}
From Lemma \ref{connected}, the free energy ${\cal E}(M^i_V)$ for each component $M^i$ can be computed
from its graph representation $\hat M^i$
(when $M^i$ is
 singleton,
its free energy is defined to be
 0).
We now define the free energy of $M_V$ to be
${\cal E}(M_V)=\max_i {\cal E}(M^i_V)$.
Clearly, using the definition of ${\cal E}(M^i_V)$ and Lemma \ref{connected}, we immediately
 have:
\begin{theorem}\label{EC}
When NFA $M$ is cleaned-up, ${\cal E}(M_V)$ is computable.
\end{theorem}
This definition  of ${\cal E}(M_V)$  is valid, since, from (\ref{eqq8976}), we can easily
 show:
\begin{theorem}\label{VPP} (Variational Principle for  NFA)
When NFA $M$ is cleaned-up, we have
${\rm G}_V(A)={\cal E}(M_V),$
where $A$ is the set of all accepting runs of $M$,
 and
$V$ is a cost function that assigns a cost in $\mathbb{R}$ to a transition
in $M$.
\end{theorem}

We now consider   a regular language $L$ associated with a cost function defined earlier as
$U: \Sigma\times \Sigma \to \mathbb{R}$. Let $M$ be an NFA with a cost function $V$.
We say that $(M,V)$ {\em implements} $(L,U)$ if
$M$ accepts $L$, and, for each $w\in L$ with length at least 2
 and each accepting run $\tau$ for $w$ in $M$, we have
$(U)(w)=(V)(\tau)$  (i.e.\ the total cost on $w$ defined by $U$ is
 kept exactly
 the same as the total cost
 of
 each accepting run
$\tau$ defined by $V$).
We may ask a number of questions:

Q1. Let $M$ be an NFA that accepts $L$. Can we always find a $V$ such that
$(M,V)$ { implements} $(L,U)$?

Q2. Can we find an NFA $M$ that accepts $L$ and find a $V$ such that $(M,V)$ implements $(L,U)$?

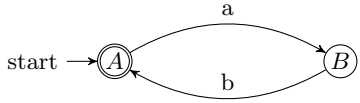
\begin{figure}
\centering
\begin{tikzpicture}[->, >=stealth', node distance=2cm, on grid, auto]
   \node[state,initial, accepting, inner sep=1pt,minimum size=0pt] (A)   {$A$};
   \node[state, inner sep=1pt,minimum size=0pt] (B) [right=3cm of A] {$B$};
   \path[->]
   (A) edge [bend left] node[sloped, above] {a} (B)

   (B) edge [bend left] node[above] {b}  (A);

\end{tikzpicture}

\caption{The finite automaton for Example \ref{example6}.}
\label{fig1}
\end{figure}

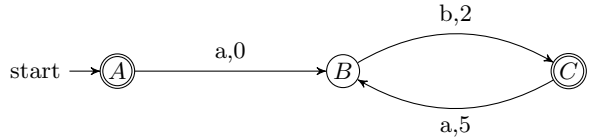
\begin{figure}
\centering
\begin{tikzpicture}[->, >=stealth', node distance=2cm, on grid, auto]
   \node[state,initial, accepting, inner sep=1pt,minimum size=0pt] (A)   {$A$};
   \node[state, inner sep=1pt,minimum size=0pt] (B) [right=3cm of A] {$B$};
   \node[state,accepting, inner sep=1pt,minimum size=0pt] (C) [right=3cm of B] {$C$};
   \path[->]
   (A) edge node[sloped, above] {a,0} (B)
   (B) edge [bend left] node[above] {b,2}  (C)
   (C) edge [bend left] node {a,5} (B);

\end{tikzpicture}

\caption{The finite automaton for Example \ref{example7}.}
\label{fig2}
\end{figure}

\noindent Q1 has a negative answer, unfortunately, shown in the following example.
\begin{example}
\label{example6}
 Consider $L=(ab)^*$ with $U(a,b)=2, U(b,a)=5$. Clearly, $M$ shown in Figure~\ref{fig1} accepts $L$.
However, there are no $V_a$ and $V_b$ (costs assigned on the $a$-transition and the $b$-transition in $M$)
such that the resulting $(M,V)$ implements $(L,U)$.
\end{example}
 Because of
the negative answer to Q1,  the positive answer to Q2  is meaningful.
\begin{example} \label{example7}
Following the previous example,
we can draw an $M'$ shown in Figure \ref{fig2} (the costs $V$ are assigned to the transitions in the figure). One can verify
that $(M,V)$ implements $(L,U)$.
\end{example}

\begin{theorem}\label{eqq0212}
For each regular language $L$ associated with a cost function $U$:
(1). For each NFA $M$ and cost function $U$ for transitions in $M$ such that
$(M,V)$ implements $(L,U)$, we have
${\rm G}_U(L)\le {\cal E}(M_V).$
(2).   There is a DFA $M$ and cost function $U$ for transitions in $M$ such that
$(M,V)$ implements $(L,U)$.  Furthermore, we have
${\rm G}_U(L)={\cal E}(M_V).$
(3). When $L$ is prefix closed, the limit (instead of limsup)  in the definition of ${\rm G}_U(L)$ in formula (\ref{eqq9004}) exists.
\end{theorem}

It is difficult to show whether a nonregular language $L$ associated with a cost function
$U$ has a computable free energy (see also Theorem \ref{diff}).
 Below, we show a class of languages where the energy is computable.

We first need a definition.
Let $Y$ be a finite set of  integer variables.  An atomic Presburger formula
on $Y$ is either a linear constraint
$\sum_{y\in Y} a_yy<b$, or   a mod constraint $x\equiv_d c$,
where $a_y, b, c$ and $d$ are integers with $0\le
 c \le d$.
A Presburger formula can always be constructed from atomic Presburger formulas
using $\neg$ and $\land$.  Presburger formulas are closed under quantification.
It is well-known that it is decidable whether a Presburger formula is satisfiable.

Let $P(\theta_1,\cdots,\theta_k)$ be a Presburger formula over nonnegative integer variables $\theta_1,\cdots,\theta_k$, for some $k$.
For each $1\le i\le k$, we associate
 $P$ with a regular language  $r_i$. We use ${\bf r}$ to denote $\langle r_1,\cdots, r_k\rangle$.
We then define a language $L_{P,{\bf r}}$ as the set of all words in the form of
$w_1\cdots   w_k$
such that each $w_i\in r_i$, and the lengths $|w_i|$ of $w_i$'s satisfy
$P(|w_1|,\cdots, |w_k|)$.  A {\em linear-length} language  $L$ is specified by
a regular language $L'$ along with $P$ and ${\bf r}$ such that
$L=L'\cap L_{P,{\bf r}}$.
Intuitively,
 $L$ is a subset of given regular language such that each word in the subset
is the concatenation of $k$  subwords, each of which is
drawn from a regular language and the lengths of the subwords are constrained by a Presburger formula. For instance,
$L=\{a^nb^{2n}a^{3n}: n\ge 0\}$ is a linear-length language, which is not context-free.
We can show:

\begin{theorem}\label{linearl}
For linear-length language $L$ and cost function $U$,
the free energy  ${\rm G}_{U}(L)$ is computable (from $L$'s specification and
$U$).
The computability remains even when $L$ is a finite union of linear-length languages.
\end{theorem}

Suppose that symbols in $\Sigma$ are $a_1,\cdots, a_k$, for some $k$.
Let $P(\theta_1,\cdots,\theta_k)$ be a Presburger formula over nonnegative integer variables $\theta_1,\cdots,\theta_k$.
Recall that we use $\#_{a_i}(\alpha)$ to denote the number of appearances of symbol $a_i$
 in
 word $\alpha$.
For a given regular language $L$, we use $L_P$ to denote the set of all words $\alpha$ in $L$ such that
$P(\#_{a_1}(\alpha),\cdots,\#_{a_k}(\alpha))$ holds.  Such a language is called a counting language in \cite{CuiDFI14}.
Notice that the aforementioned linear-length language can be converted into a counting language
through
 proper
symbol renaming. In \cite{CuiDFI14}, it was shown that the information rate of  counting languages
is computable using a complicated convex-optimization approach. Currently, we do not know if a similar technique can be
used to show
 that
the free energy of a counting language is computable.

In a linear-length language, we put a Presburger constraint on the lengths of certain subwords. Now, we study the case where we put
a  similar constraint on the total costs of the subwords.
Let $P(\theta_1,\cdots,\theta_k)$ be a formula in additive theory of rationales
over rational variables $\theta_1,\cdots,\theta_k$, for some $k$. Recall that
 ${\bf r}$ is an array of $k$ regular languages
$r_1,\cdots,r_k$. Let $U$ be a cost function.
We  define a language $L_{P,{\bf r},U}$ as the set of all words in the form of
$w_1\cdots   w_k$
such
 that
each $w_i\in r_i$, and the total costs $c_i=( U)(w_i)$ on $w_i$'s satisfy
$P(c_1,\cdots, c_k)$.  A {\em linear-cost} language  $L$ is specified by
a regular language $L'$ along with $P$, ${\bf r}$ and $U$ such that
$L=L'\cap L_{P,{\bf r},U}$.

\begin{example}\label{ex09}
Consider a word $w$ in the form of
$(a+b)^*(c+d)^*(a+b)^*$, with cost function $U$ given. Conceptually,
 $w$
is
 rewritten as
$w_1w_2w_3$ with $w_1\in r_1=(a+b)^*$,
$w_2\in r_2=(c+d)^*$ and $w_3\in r_3=(a+b)^*$.
We use $c_1, c_2, c_3$ to
 denote the total costs on
$w_1, w_2$ and $w_3$, respectively, and
 add
all such $w$ satisfying $c_1>c_2>c_3>0$ into language $L$.
 Then
$L$ is a linear-cost language. Clearly,
 $L$ is not context-free in general, even when
the cost function is nonnegative.
\end{example}

Currently, we do not know whether the free energy
 of a linear-cost language is computable or not.
Even for the following simple $L$, the problem  is open:  the set $L$
 consists
of all $w$ in a given regular language $L'$
such that the total cost on $w$ (with respect to a given cost function) is zero.  We believe that
for such a simple $L$, the energy is computable by generalizing the seminal proof by Kuich in \cite{Kuich70}.

Even when the cost function is nonnegative,  the problem is open.
However, for nonnegative cost function, we have a computable special case.
 A restricted linear-cost language is a linear-cost language where the formula
$P(\theta_1,\cdots,\theta_k)$ in defining the language takes the following special form: it is a disjunction over
conjunctions of
\begin{equation}\label{eqq1209}
k_i\cdot \theta_i \sim k_j\cdot \theta_j+ l
\end{equation}
where $\sim\in\{>.<.=,\ge,\le\}$, and for each $i$,
 a nonnegative integer $k_i$
is unique throughout a conjunction, while
$l$ is only tied with the formula in (\ref{eqq1209}).
For instance,  $(2\theta_1>3\theta_2-5 \land 2\theta_1=4\theta_3-8)\lor (7\theta_1<3\theta_2-12 \land
7\theta_1>3\theta_2+17)$ is a valid formula, but
$(2\theta_1>3\theta_2-5 \land 5\theta_1=4\theta_3-8)\lor (7\theta_1<3\theta_2-12 \land
7\theta_1>3\theta_2+17)$ is not.
A restricted linear-cost language is not  necessarily context-free in general as shown in  Example \ref{ex09}.
(The following result still holds when we replace each term $k_ix_i$ with a nonnegative integer  linear combination
of a subset $V_i$ of variables $\theta_1,\cdots,\theta_k$, as long as the combination is unique throughout a conjunction
and the sets $V_i$ are disjoint.  We omit the details.)
\begin{theorem} \label{theorem11}
For restricted linear-cost language $L$ and nonnegative cost function $U$,
 the free energy  ${\rm G}_{U}(L)$ is computable (from $L$'s specification and
 $U$). The computability remains even when $L$ is a finite union of
restricted linear-cost
 languages.
\end{theorem}

We turn back to NFA $M$ (which is cleaned-up) with a cost function $V$ on transitions.
 We
show an application of the free energy of
$(M,V)$ in estimating nondeterminism in an NFA.
There are  applications for
 such an
estimation.  In software engineering, NFAs can be used to specify
a highly nondeterministic system such as a concurrent system where the input symbols are the observable events.
Being nondeterministic, the same sequence of input symbols can have many different execution sequences. Hence, an ideal measure
for the nondeterminism would be the asymptotic growth rate of the ratio $f(n)/g(n)$ where
$f(n)$ is the total number of executions of input sequences of length $n$ while $g(n)$ is the total number of input sequences of length $n$.
More precisely,  we define  (slightly different from the above)
$g(n)$ to be the number of words $\alpha$ of length $\le n$ in $L(M)$, and
$f(n)$ to be the number of initialized runs of all
 $\alpha$'s.
Then
the nondeterminism in $M$ is defined by
$$\lambda_M=\lim_{n\to\infty} {{\log f(n)-\log g(n)}\over n}.$$
Clearly,  the limit in $\lambda_M\ge 0$ exists and
 is
finite (since runs are
 prefix closed, and $M$ has no $\epsilon$-transitions).
In particular, when $M$ is deterministic, $\lambda_M=0$.

In reality, such a metric is relevant. For instance, it can be used to estimate a form of
 coverage of extensive testing
of a nondeterministic system (e.g.\ how many paths have already been  exercised for an average input sequence). The estimation is important since
it is well-known
 that
nondeterministic software systems are notoriously difficult and costly to test.
However, in computing
 $\lambda$, the difficult part is the asymptotic growth rate of $\log g(n)$ since currently available algorithms
 must
convert
 NFA
$M$ into a
 DFA
and use the Chomsky-Miller  algorithm \cite{ChomskyM58}. The conversion may cause an exponential blow-up
 in
the
 number of states in the DFA,
which is not tractable.  We need a practically efficient algorithm to give an estimation
of
$\lambda_M$.

We propose an efficient  estimation  approach  based on free energy.
For the given NFA $M$, we define a cost function $V$ on transitions of $M$ as follows.
Let $k(p,a)$ be the total number of $p''$ such that
$p{\stackrel{a}\to}p''$ is a transition in $M$.
For each transition
$p{\stackrel{a}\to}p'$ in $M$, we define $V(p,a,p')=\ln k(p,a)$.
Notice that if $k(p,a)=1$ (in this case, $M$ is deterministic at state $p$ on input $a$),
$V(p,a,p')=0$. Otherwise (i.e., $k(p,a)>1$. In this case, $M$ is nondeterministic at state $p$ on input $a$),
$V(p,a,p')=\ln k(p,a)>0$.

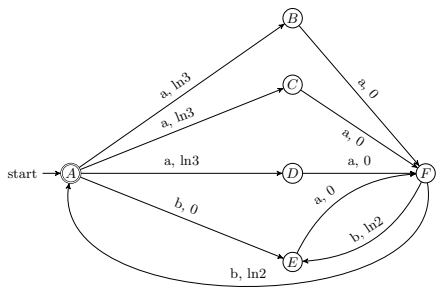
\begin{figure}
\centering
\resizebox{0.5\textwidth}{!}{
\begin{tikzpicture}[->, >=stealth', node distance=2cm, on grid, auto]
   \node[state,initial, accepting, inner sep=1pt,minimum size=0pt] (A)   {$A$};
   \node[state, inner sep=1pt,minimum size=0pt] (D) [right=5cm of A] {$D$};
   \node[state, inner sep=1pt,minimum size=0pt] (B) [above=3.5cm of D] {$B$};
   \node[state, inner sep=1pt,minimum size=0pt] (C) [above=2cm of D] {$C$};

   \node[state, inner sep=1pt,minimum size=0pt] (E) [below=2cm of D] {$E$};
   \node[state, inner sep=1pt,minimum size=0pt] (F) [right=3cm of D] {$F$};
   \path[->]
   (A) edge node[sloped, above] {a, ln{3}} (B)
       edge node[sloped, above] {a, ln{3}} (C)
       edge node[sloped, above] {a, ln{3}} (D)
       edge node[sloped, above] {b, 0} (E)
   (B) edge             node[sloped, above] {a, 0}  (F)
   (C) edge             node[sloped, below] {a, 0}  (F)
   (D) edge             node {a, 0}  (F)
   (E) edge [bend left] node[sloped] {a, 0}  (F)
   (F) edge [bend left] node[sloped, above] {b, ln{2}}  (E)
       edge [bend left=100] node[above] {b, ln{2}}  (A);
\end{tikzpicture}
}
\caption{The finite automaton for Example \ref{example12} with costs labeled.}
\label{fig3}
\end{figure}

\begin{example} \label{example12}
Figure \ref{fig3} shows an NFA with the $V$ assigned on transitions.
\end{example}

We now use  the free energy difference  $\lambda^+_M={\cal E}(M_V)-{\cal E}(M_{\bf 0})$ to estimate $\lambda_M$. Herein,
$M_{\bf 0}$ is the $M$ where each transition is assigned with cost 0.
Notice that $\lambda^+_M\ge 0$
(roughly, from the Perron–Frobenius theorem applied on the Gurevich matrices for $M_V$ and for $M_{\bf 0}$)
and,
when $M$ is deterministic,  $\lambda^+_M=\lambda_M=0$.

We first intuitively explain the meaning behind $\lambda^+_M$.  In $M_V$, each transition $p{\stackrel{a}\to}p'$ is assigned
 a cost
which is the number of nats (information units in natural  logarithm)
needed to code the transition when $p$ and $a$ are given. Hence, the total cost of an average (initialized) run on a word $\alpha$
will be the total number of nats to code the run (which starts from the known initial state) when
 $\alpha$ is known.
This total cost divided by the length $n$ of the word $\alpha$ will result in nat rate $\delta$  of the code. Notice that, in the definition of
free energy of $M$, there are two parts:  the average cost per step (which roughly corresponds to the nat rate $\delta$)
and the metric entropy (which roughly corresponds to the ``average'' natural  logarithm of the branching factor at a state in $M$).
Now, in $M_{\bf 0}$, the free energy is the ``average'' natural  logarithm of the branching factor at a state in $M$.
Hence, the $\lambda^+_M={\cal E}(M_V)-{\cal E}(M_{\bf 0})$ is roughly equal
 to
the nat rate $\delta$ (to encode the transition per step for a given
input), which is the intended meaning in $\lambda_M$. When $M$ is deterministic, the input $\alpha$ decides the run and hence the nat rate is of course zero
(no extra nat is needed to encode the run).


We now compute the estimation $\lambda^+_M$ for the example NFA $M$
in Figure \ref{fig3} and obtain $\lambda^+_M={\cal E}(M_V)-{\cal E}(M_{\bf 0})=1.0850-0.5857=0.4993$.
Next, we convert the NFA into a DFA (using an online tool) $M'$ and compute
$\lambda_M={\cal E}(M_{\bf 0})-{\cal E}(M'_{\bf 0})=0.5857-0.3603=0.2255$. Indeed,
$\lambda^+_M$ is an upper estimation of $\lambda_M$, as shown below.

\begin{theorem}\label{thmupper}
 For a cleaned-up NFA $M$,
$\lambda^+_M\ge\lambda_M$.
\end{theorem}


We
 note that in Theorem \ref{thmupper},  the
upper estimation $\lambda^+_M$  can be efficiently computed from $M$.
  Finally, we point out that the estimation is asymptotically tight; the proof will be included in the full version of the paper.
We shall also point out all the results can be generalized to cost function $U$ over $k>2$ ($k$ is constant) symbols
instead of two symbols.

\section{A free-energy based similarity metric for automata and languages}
 We are now ready to use the theory developed so far to define the similarity metric.
Let $M^i$ ($i=1,2$) be an NFA with a cost function $V_i$ assigned on edges.
Notice that the two automata $M^1$ and $M^2$ work on the same input alphabet $\Sigma$.
Consider a word $w=a_0\cdots a_{n-1}$ accepted by both automata.
Suppose that the following transition sequences
$(q_0,a_0,q_1)(q_1,a_1,q_2)\cdots (q_{n-1},a_{n-1},q_n)$ in $M^1$
and
$(p_0,a_0,p_1)(p_1,a_1,p_2)\cdots (p_{n-1},a_{n-1},p_n)$ in $M^2$
witness the acceptance.
However, the total cost (i.e., the total energy or potential) on the first accepting sequence is
defined by $V_1(q_0,a_0,q_1)+\cdots+V_1(q_{n-1},a_{n-1},q_n)$ while the
total cost on the second is
$V_2(p_0,a_0,p_1)+\cdots+V_2 (p_{n-1},a_{n-1},p_n)$.
The sum of the two costs shall tell the deviation of the free energy on the two sequences
on the input word $w$.
The sum can be expressed on each individual transition as
$V_1(q_0,a_0,q_1)+V_2(p_0,a_0,p_1)+\cdots+V_1(q_{n-1},a_{n-1},q_n)+V_2 (p_{n-1},a_{n-1},p_n)$,
which again is the free energy on a properly defined (in below) ``shared sequence" between the two automata.

We define $M$ to be the Cartesian product of $M^1$ and $M^2$ in a standard way.  That is,
$((q,p), a, (q',p'))$ is a transition in $M$ iff
$(q,a,q')$ is a transition in $M^1$ and $(p,a,p')$ is a transition in $M^2$,  for all states $p,q,,p',q'$.
The initial state in $M$ is the pair of initial states in $M^1$ and in $M^2$ and
the accepting states in $M$ are all the pairs of an accepting state in $M^1$ and an accepting state in $M^2$.
Again, we assume that $M$ is cleaned up. Clearly, $M$ is an NFA accepting $L(M^1)\cap L(M^2)$.

We now define  the cost functions $V$  on the $M$ as
$V((q,p), a, (q',p'))=V_1(q,a,q')+V_2(p,a,p')$. The semantic similarity metric
$\Delta(M^1_{V_1},M^2_{V_2})$ is defined as
the free energy of $M$; i.e., ${\cal E}(M_V)$. Intuitively, this definition catches the
average ``shared free energy" per step on the shared accepting runs between $M_1$ and $M_2$.
We have
\begin{equation}\label{eqq0191}
0\le \Delta(M^1_{V_1},M^2_{V_2}) \le {\cal E}(M^1_{V_1})+{\cal E}(M^2_{V_2}).
\end{equation}
To see (\ref{eqq0191}),   $\Delta(M^1_{V_1},M^2_{V_2})={\cal E}(M_V)\ge 0$ is obvious,  since
 the LHS of
(\ref{eqq8976}) in Theorem \ref{NFAVP}  is nonnegative and so is the LHS of the equation in Theorem \ref{VPP}.
To show $\Delta(M^1_{V_1},M^2_{V_2}) \le {\cal E}(M^1_{V_1})+{\cal E}(M^2_{V_2})$ in
(\ref{eqq0191}), we need some effort.
First we assume that $M$ is strongly connected and hence we can use
Theorem \ref{NFAVP}.  Observe that
 the term
$\sum_{\tau\in A, |\tau|=n}
  e^{(V)(\tau)}$ in
(\ref{eqq8976}) in Theorem \ref{NFAVP}
satisfies, using the definition $V$,
 the following inequality
$$\sum_{\tau\in A, |\tau|=n}
  e^{(V)(\tau)} \le \sum_{\tau_1\in A_1, |\tau_1|=n}
  e^{(V_1)(\tau_1)} \cdot \sum_{\tau_2\in A_2, |\tau_2|=n}
  e^{(V_2)(\tau_2)}$$
where $A_1$ and $A_2$ are accepting transition sequences of $M_1$ and $M_2$, respectively.
Now we plug-in the RHS of   the inequality into (\ref{eqq8976})  and obtain
$$\limsup_{n\to\infty}{1\over n}\ln \sum_{\tau\in A, |\tau|=n}
  e^{(V)(\tau)} \le \limsup_{n\to\infty}{1\over n}\ln \sum_{\tau_1\in A_1, |\tau_1|=n}
  e^{(V_1)(\tau_1)}$$
$$+  \limsup_{n\to\infty}{1\over n}\ln \sum_{\tau_2\in A_2, |\tau_2|=n}
  e^{(V_2)(\tau_2)}.$$
Then, we use Theorem \ref{VPP} on $M^1_{V_1}$ and $M^2_{V_2}$ and hence the RHS of the above inequality becomes
$$\limsup_{n\to\infty}{1\over n}\ln \sum_{\tau\in A, |\tau|=n}
  e^{(V)(\tau)} \le  {\cal E}(M^1_{V_1})+{\cal E}(M^2_{V_2}).$$
Again, using (\ref{eqq8976})  in Theorem \ref{NFAVP} on the LHS of the inequality, we have
 $${\cal E}(M_V)\le {\cal E}(M^1_{V_1})+{\cal E}(M^2_{V_2}).$$
Using Theorem \ref{VPP} again on the LHS, we finally obtain for a general $M$ (which may not be strongly connected),
the above inequality still holds, which is essentially the result in (\ref{eqq0191}).

In particular,  the reader can easily check that the inequality in (\ref{eqq0191}) is tight:
when $M^1$ and $M^2$ are completely independent (i.e., $L(M^1)\cap L(M^2)=\emptyset$),
the similarity metric $ \Delta(M^1_{V_1},M^2_{V_2})=0$.
However, when $M^1$ and $M^2$ are the same and the $V_1$ and $V_2$ are also the same,
we have the similarity metric $ \Delta(M^1_{V_1},M^2_{V_2})$
reaches the maximum  ${\cal E}(M^1_{V_1})+{\cal E}(M^2_{V_2})$. Intuitively,
the metric $ \Delta(M^1_{V_1},M^2_{V_2})$ characterizes the ``shared free energy" between
the two finite state programs $M^1_{V_1}$ and $M^2_{V_2}$ as follows.
Imagine that the two programs (automata) are two moving gas molecules. At each step of observation,
the molecules can be highly random and hence it makes little sense to compare
every step of the observations to figure out  the similarity between the two molecules.
The approach we take in defining the metric $ \Delta(M^1_{V_1},M^2_{V_2})$ is to ``create" a third
molecule $M_V$ that, at each step, possesses the potential as the sum of the first two molecules (this is a reward)
whenever the first two molecule share the same orbit (i.e., the same input  symbol)  -- otherwise
when the orbit are different, the potential of the third molecule is $-\infty$ (this is a penalty).
Clearly, the dynamics of the third molecule would be very different from the first two.  However, as we
have shown above, the long term characteristic of  free energy of the third molecule reflects the shared free
energy between the first two molecules when they follow the same orbit.

\begin{figure}
\centering
\resizebox{0.8\textwidth}{!}{
\begin{tikzpicture}[->, >=stealth', node distance=2cm, on grid, auto]
   \node[state,initial, inner sep=1pt,minimum size=0pt] (1)   {$1$};
   \node[state, accepting, inner sep=1pt,minimum size=0pt] (2) [right=3cm of 1] {$2$};
   \node[state, inner sep=1pt,minimum size=0pt] (4) [below=3cm of 2] {$4$};
   \node[state, accepting, inner sep=1pt,minimum size=0pt] (3) [left=3cm of 4] {$3$};
   \path[->]
   (1) edge node[above] {A, 0.14} (2)
       edge node[below] {G, 0.1} (2)
       edge  [bend right]  node[sloped, below] {A, 0.76} (3)
   (2) edge  [bend right]              node[sloped, above] {G, 0.2}  (1)
       edge          node[sloped, above] {T, 0.8}  (4)
   (4) edge [bend left] node {C, 0.35 } (2)
       edge             node {C, 0.65}  (3)
   (3) edge             node[sloped, below] {G, 0.7}  (1);
\end{tikzpicture}

\begin{tikzpicture}[->, >=stealth', node distance=2cm, on grid, auto]
\node[state, initial, accepting, inner sep=1pt, minimum size=0pt] (5) {$5$};
\node[state, accepting, inner sep=1pt,minimum size=0pt] (6) [right=3cm of 5] {$6$};
\path[->]
(5) edge         node[above] {C, 0.2; G, 0.4} (6)
    edge         node[below] {A, 0.9} (6)
(6) edge [bend right=100] node[sloped, above] {T, 0.7} (5)
   edge [bend left=100 ] node[sloped, below] {C, 0.2} (5);
\end{tikzpicture}
\begin{tikzpicture}[->, >=stealth', node distance=2cm, on grid, auto]
   \node[state,initial, inner sep=1pt,minimum size=0pt] (15)   {$1,5$};
   \node[state, accepting, inner sep=1pt,minimum size=0pt] (26) [right=3cm of 15] {$2,6$};
   \node[state, inner sep=1pt,minimum size=0pt] (45) [right=3cm of 26] {$4,5$};
   \node[state, accepting, inner sep=1pt,minimum size=0pt] (36) [below=1.5cm of 26] {$3,6$};
   \path[->]
   (15) edge node {G, 0.5; A, 1.04} (26)
        edge [bend right] node{A, 1.66} (36)
   (26) edge [bend left] node {T, 1.5}  (45)
   (45) edge   node {C, 0.55 } (26)
       edge              node {C, 0.85}  (36);
\end{tikzpicture}
}

\caption{The first figure is an NFA $M^1_{V_1}$ with cost function $V_1$;
the second figure is an NFA $M^2_{V_2}$ with cost function $V_2$;
the third figure is the Cartesian product $M_V$.}
\label{fig1001}
\end{figure}
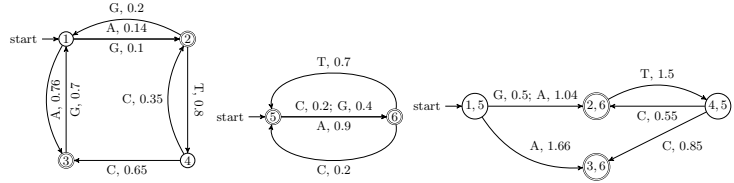

We now look at an example of two NFAs  $M^1_{V_1}$ and $M^2_{V_2}$ where the cost functions $V_1$ and
$V_2$ are labelled in the Figures \ref{fig1001}.
One can think that the two automata try to specify a genome pattern on the  four nucleotides
($G,A,C,T$) in DNA while the costs could be interpreted as probabilities, weights of choices, etc.
Notice that the semantics of the automata are the nucleic acid sequences that the automata accept, and those
sequences are associated with a cost on each nucleotide.
For instance, the following sequence $(A, 0.14)(T,0.8)(C,0.65)$ is accepted by $M^1_{V_1}$.
Hence, the semantic similarity between the two automata shall  concern the similarity between the sequences
(with costs) accepted by the two automata instead of the naive similarity on the appearances of the
two graphs themselves. Again, we imagine each such sequence accepted as a molecule moving along the orbit
specified on the sequence and use the shared free energy between the molecule specified by $M^1_{V_1}$
and the molecule specified by $M^2_{V_2}$ to measure the semantic similarity
$\Delta(M^1_{V_1},M^2_{V_2})$. We compute, using the algorithms in Theorem \ref{VPP} and the definition
of the Cartesian product $M_V$ shown earlier, the results
${\cal E}(M^1_{V_1})=0.3500,  {\cal E}(M^2_{V_2})=1.4087, $
and the similarity metric $\Delta(M^1_{V_1},M^2_{V_2})=1.025$, which indeed satisfies (\ref{eqq0191}).

 Notice that $\Delta(M^1_{V_1},M^2_{V_2})$ can be computed efficiently (which involves only Cartesian product of
the two automata, and largest eigenvalues of the Gurevich matrices in Theorems \ref{VPP}  and \ref{NFAVP}).

Let $L_1$ and $L_2$ be two regular languages associated with two cost functions $U_1$ and $U_2$, respectively,
According to  Theorem \ref{eqq0212} (2),  we can construct  DFAs $M^1_{V_1}$ and $M^2_{V_2}$ and use
$\Delta(M^1_{V_1},M^2_{V_2})$ to serve as the semantic similarity metric
for the two regular languages. However, it does not seem that $\Delta(M^1_{V_1},M^2_{V_2})$ can be efficiently computed from the regular languages since the known construction from regular languages to deterministic
finite automata involves exponential blowup on the state space.
There might be other alternative definitions on the metric over regular languages
(such as using the estimation of nondeterminism in an NFA shown earlier in the paper) such that the metric can be
efficiently computed. We leave this for future work.

\section*{Acknowledgements}
We would like to thank  Jean-Charles Delvenne, David Koslicki, Daniel J. Thompson, Eric Wang,  William
J. Hutton III,  and Ali Saberi for  discussions.

\bibliography{mybib}{}
\bibliographystyle{plain}

\newpage

\appendix
\section{Appendix --  proofs}

\textbf{Proof of Theorem \ref{diff}}

\begin{proof}Let $L$ be a context-free language on alphabet $\Sigma=\{a,b\}$.
Consider the language $L'=(\Sigma^*-L)\Sigma^*$ and $U$ to be the zero function.
It is left to the reader to check that, if $L$ is total then
$G_U(L')=0$ otherwise $G_U(L')=\ln 2$. The result follows since totalness of
context-free languages is well-known undecidable.
\end{proof}

\textbf{Proof of Theorem \ref{NFAVP} }

\begin{proof} Let ${\bf M}$ be the Gurevich matrix of $M$.
Since $M$ is strongly connected, ${\bf M}$ is irreducible.
From the definition, ${\bf M}$ is nonnegative with $\lambda>0$ being its
Perron-Frobenius eigenvalue. From the results in \cite{Gurevich1984} (and also
the Perron-Frobenius theorem), let $\eta$ (a row vector) and $\xi$ (a column vector)
be the unique positive left eigenvector and the unique positive
right eigenvector of ${\bf M}$, respectively, of the eigenvalue $\lambda$,
satisfying
the normalization condition:  $\eta\xi=1$.  (A vector is positive if all its elements are positive reals.)

We now show (\ref{eqq8977}), where we use $W(n)$ to denote
$\sum_{
\tau\in R, |\tau|=n}
 e^{(V)(\tau)}$.
Observe that, for each $n$, by definition,
\begin{equation}\label{ex90}
W(n)+W(n+1)=\| {\bf M}^{2n}\|
\end{equation}
 that is the sum of all elements in matrix ${\bf M}^{2n}$.
(The term of $2n$ comes from the fact that the Gurevich matrix is constructed from $\hat M$. In $\hat M$, by definition,
each transition in $M$ is translated into two transitions in $\hat M$.)
Let $g_{2n}(j)$ be the $j$-th column sum of  ${\bf M}^{2n}$.
Recall that ${\bf M}$ is a $k\times k$ matrix and therefore
\begin{equation}\label{ex91}
\| {\bf M}^{2n}\|=g_{2n}(1)+\cdots+ g_{2n}(k).
\end{equation}
  Using the fact that $\xi=(\xi_1,\cdots,\xi_k)$ is a
right eigenvector, we have ${\bf M}^{2n}\xi=\lambda^{2n}\xi$.  This gives
\begin{equation}\label{ex92}
\xi_1g_{2n}(1)+ \cdots + \xi_k g_{2n}(k)=
\lambda^{2n}(\xi_1+\cdots+\xi_k).
\end{equation}
 Since $\xi$ is a positive vector,
both $\xi_{\min}=\min(\xi_1,\cdots,\xi_k)$ and $\xi_{\max}=\max(\xi_1,\cdots,\xi_k)$ are positive.
Therefore, from (\ref{ex92}) and (\ref{ex91}), we have
$   \xi_{\min}  \| {\bf M}^{2n}\|   \le \lambda^{2n}(\xi_1+\cdots+\xi_k)\le \xi_{\max}\| {\bf M}^{2n}\|$.
Finally, combining (\ref{ex90}), we have, for each $n$,
\begin{equation}\label{ex93}
  c_0\lambda^{2n}  \le W(n)+W(n+1)  \le  c_1\lambda^{2n},
\end{equation}
  where the positive constants $c_0$ and $c_1$ only depends on
$\xi$ (not on $n$).  Hence,  the LHS of (\ref{eqq8977}),  ${\rm G}_V(R)$, by definition,
is
$\limsup_{n\to\infty} {1\over n}\ln W(n)\le \limsup_{n\to\infty} {1\over n}\ln (W(n)+W(n+1))\le
2\ln\lambda
$, using (\ref{ex93}). On the other hand,
from (\ref{ex93}), one of $W(n)$ and $W(n+1)$ must be at least
${1\over 2}c_0\lambda^{2n}$.  This gives $\liminf_{n\to\infty} {1\over n}\ln W(n)\ge 2\ln\lambda$.
Hence, the limit in  (\ref{eqq8977}) exists and equals ${\cal E}(M_V)=2\ln\lambda$.

To show (\ref{eqq8976}), let $$Z(n)=\sum_{
\tau\in A, |\tau|=n}
 e^{(V)(\tau)}.$$
By definition,
${\rm G}_V(A)=\limsup_{n\to \infty} {1\over n} \ln Z(n) $.
However, to prove that the limsup is indeed equal to $2\ln\lambda$, we take two steps.
First notice that ${\rm G}_V(A)\le {\rm G}_V(R)$ since $A\subseteq R$. This gives, from
 (\ref{eqq8977}),   ${\rm G}_V(A)\le 2\ln\lambda$.  It is left to us to show the limsup is at least $2\ln\lambda$.
Observe that for any fixed constant $N_0$, we have
\begin{equation}\label{ex61}
\limsup_{n\to \infty} {1\over n} \ln Z(n)=\limsup_{n\to \infty} {1\over n} \ln (Z(n)+\cdots+Z(n+N_0)),
\end{equation}
using the converging subsequence definition of limsup.
Now, take $n_0$ to be the largest of the lengths of shortest paths (in terms of the number of transitions on a
 path)  from the initial state  to each individual state in $M$ and from
each individual state in $M$ to each accepting state in $M$.
For each of the  shortest paths, we obtain the total cost on the path (which is the sum of costs $V(t)$ on the transitions
$t$ in $M$ on the path). Let $\epsilon$ be the minimal of all such total costs.
Let $N_0=2\cdot n_0$. Notice that both $\epsilon$ and $N_0$ are constants (not depending on $n$).
We can show $Z(n)+\cdots+Z(n+N_0)\ge e^\epsilon \cdot W(n)\cdot e^\epsilon$ where $W(n)$ is defined in the proof above.
This is because every run of length $n$ in $R$ is a substring of an accepting run with length at most $n_0+n+n_0=N_0+n$.
Hence, using (\ref{ex61}) and (\ref{eqq8977}) and noticing that
${\cal E}(M_V)=2\ln\lambda$,  we have $\limsup_{n\to \infty} {1\over n} \ln Z(n)\ge 2\ln\lambda$.

We now show (\ref{eqq8975}), under the assumption that every state in $M$ is an accepting state. That is, under the assumption,
the limit of
$\lim_{n\to \infty} {1\over n} \ln Z(n) $ exists. From (\ref{eqq8976}), we only need show
\begin{equation}\label{ex24}
\liminf_{n\to \infty} {1\over n} \ln Z(n)\ge  2\ln\lambda.
\end{equation} Now, take $n_0$ to be the largest of
the lengths of shortest paths  from the initial state  to each individual state in $M$.
For each run of length not more than $n_0$ (which not necessarily start from the initial state),
we obtain the total cost on the path. Let $\epsilon$ be the minimal of all such total costs.
Observe that
\begin{equation}\label{ex23}
Z(n+n_0)\ge \min(e^\epsilon \cdot W(n), e^\epsilon \cdot W(n)\cdot e^\epsilon).
\end{equation}
This is because every run $\tau$ of length $n$ in $R$ is a substring of an accepting run $\alpha$ with length $n+n_0$.
To see this,  suppose that the $\tau$ starts with $p$ ends with $q$. If the shortest run $\beta$
from the initial state to the $p$ is with length
$l=n_0$ already, then simply take $\alpha$ to be the shortest run $\beta$ concatenated with $\tau$. In this case,
the total cost on the $\alpha$ is at least $\epsilon+ (V)(\tau)$.
On the other hand, if the length of $\beta$ is $l<n_0$, then arbitrarily choose a run $\gamma$
from $q$ with length $n_0-l$. In this case, we take $\alpha$ to be the $\beta$ concatenated with $\tau$ and then $\gamma$.
Clearly, the total cost on the $\alpha$ is at least $\epsilon+ (V)(\tau)+ \epsilon$. This gives the inequality in
 (\ref{ex23}), using the definitions of
$Z(\cdot)$ and $W(\cdot)$.
From (\ref{ex23}), the result in (\ref{ex24}) is immediate, using  (\ref{eqq8977}) and noticing that
${\cal E}(M_V)=2\ln\lambda$.
\end{proof}

\textbf{Proof of Theorem \ref{eqq0212}}

\begin{proof}
(1).  The result directly follows from the definition of ${\rm G}_U(L)$ in (\ref{eqq9004}), the variational principle of NFA
in Theorem \ref{VPP}, and the fact that the mapping from a word in $L$ to an accepting run of $w$ in $M$ is many-to-one.

(2). Let $M'$ be a DFA accepting $L$.  We now construct
 the desired DFA $M$ as follows.   $(p,a_1,q){\stackrel{a_2}{\to}}  (q,a_2,s)$ is a transition in $M$ (with cost $V$ on this transition being
$U(a_1,a_2)$)
 iff
$p{\stackrel{a_1}{\to}} q$ and $q{\stackrel{a_2}{\to}} s$ are transitions in $M'$.
In addition to this, $p_0$ is also the initial state in $M$ with transitions
$p_0{\stackrel{a}{\to}}(p_0,a,p)$ (with cost $V$ on this transition being 0)
  where it is the initial state in $M'$ and $p_0{\stackrel{a}{\to}} p$ is a transition in $M'$.
In particular,
Each $(q,a,s)$ is an accepting state in $M$ if $s$ is an accepting state in $M'$. Clearly,
$(M,V)$ implements $(L,U)$. Notice that the mapping from a word in $L$ to an accepting run of $w$ in $M$ is one-to-one.
Hence, we have the equation in Theorem \ref{eqq0212} (2), from
the definition of ${\rm G}_U(L)$ in (\ref{eqq9004}), the variational principle of NFA
in Theorem \ref{VPP}.

(3). Notice that when $L$ is prefix closed, the DFA constructed above has this property: every state is an accepting state.
The result follows from the above proof and the equation (\ref{eqq8975}) in Theorem \ref{NFAVP}.
\end{proof}

\textbf{Proof of Theorem \ref{linearl}}

\begin{proof}
We first need some well-known facts.
Let $\mathbb N$ be the set of nonnegative integers.
${\bf D}\subseteq {\mathbb N}^k$ is a {\em linear set}
 if ${\bf D}=\{{\bf d}: {\bf d}={\bf d}_0+s_1{\bf d}_1+\cdots + s_i{\bf d}_m,
 s_1,\cdots,s_m\ge 0\}$
where ${\bf d}_0,\cdots,{\bf d}_m$ are constant $k$-arity vectors in ${\mathbb N}^k$, for some
$m\ge 0$. ${\bf D}$ is a {\em semilinear set} if it is the union of finitely many linear sets.
 Every finite subset of ${\mathbb N}^k$ is
semilinear -- it is a finite union of linear sets whose generators
are constant vectors.  Clearly, semilinear sets are closed under
union and projection.  It is also known that semilinear
sets are closed under intersection and complementation.

Let $S$ be a set of $k$-tuples in ${\mathbb N}^k$.  $S$ is Presburger definable if
there is a Presburger formula $P(y_1,\cdots,y_k)$ such that
the set of non-negative integer solutions is exactly $S$.
It is well-known that $S$ is a semi-linear set iff $S$ is Presburger definable.
 It can be shown that ${\bf D}\subseteq {\mathbb N}^k$ is a semilinear set iff ${\bf D}$ is Presburger definable.

Let $\Sigma=\{a_1,\cdots,a_k\}$ be an alphabet. For each word
$\alpha\in\Sigma^*$,  define the Parikh map of $\alpha$ to be
the vector $\#(\alpha)=(\#_{a_1}(\alpha),\cdots, \#_{a_k}(\alpha))$, where
each symbol count $\#_{a_i}(\alpha)$ denotes the number of symbol $a_i$'s in
$\alpha$.
For a language $L\subseteq \Sigma^*$,  the Parikh map of $L$ is
the set $\#(L)=\{\#(\alpha): \alpha\in L\}$.
The language $L$ is semilinear if $\#(L)$ is a semilinear set.

We only proof the case when $L$ is a linear-length language.  The case when
$L$ is a finite union of linear-length languages  can be concluded directly from the footnote below.

We now continue with the proof. By definition, $L=L'\cap L_{P,{\bf r}}$, where
$L'$ is a regular language,
$P(\theta_1,\cdots,\theta_k)$ is a Presburger formula and $\langle r_1,\cdots, r_k\rangle$
are regular languages in ${\bf r}$.
Consider the set $Q=_{\rm def} \{(\theta_1,\cdots,\theta_k)\in {\mathbb N}^k:
\theta_i=\alpha_i\in r_i, 1\le i\le k\}$.  Since every regular language is a semilinear language,
it is not hard to show that $Q$ is a semilinear set (we omit the details).  The semilinear set
definable by the Presburger formula $P(\theta_1,\cdots,\theta_k)$ is also denoted by $P$.
Now, $P\cap Q$ is a semilinear set, and WLOG, we assume
\footnote{This assumption is indeed without loss of generality since, it is straightforward to
verify, from the definition in (\ref{eqq9004}),
 ${\rm G}_U(L)=\max ({\rm G}_U(L_1,\cdots,{\rm G}_U(L_k))$ whenever $L=L_1\cup\cdots\cup L_k$, for some $k$,
for any $L$.
} that it is a linear set (instead of the union of several linear sets)
${\bf D}$ in the form of
${\bf D}=\{{\bf d}: {\bf d}={\bf d}_0+s_1{\bf d}_1+\cdots + s_i{\bf d}_m,
 s_1,\cdots,s_m\ge 0\}$
where ${\bf d}_0,\cdots,{\bf d}_m$ are constant $k$-arity vectors in ${\mathbb N}^k$, for some
$m\ge 0$. In particular, WLOG, the vectors are non-zero (i.e., there is at least one positive element
in each such vector). Furthermore, the vectors ${\bf d}_1,\cdots,{\bf d}_m$ are distinct.
To make our presentation more readable and also WLOG, we assume that the first vector
${\bf d}_0$ is positive (all elements in the vector are positive).

For each constant vector ${\bf d}_i$, we create a finite
set $[{\bf d}_i]$  of  word  tuples  ${\bf b}=(\beta_1,\cdots,\beta_k)$ such that
$|\beta_j|={\bf d}_i[j]$.
We use $\langle {\bf b}\rangle$ to denote the word $\beta_1\cdots\beta_k$.
 For instance, if ${\bf d}_i=(2,0,4)$ and $\Sigma=\{a,b\}$, then
the set $[{\bf d}_i]$ is $\Sigma^2\times \Sigma^0\times\Sigma^4$.
We now consider a sequence  $\gamma$ of word tuples
${\bf b}_0{\bf b}_1\cdots {\bf b}_l$, for some $l$, drawn from $[{\bf d}_0][{\bf d}_1]^*[{\bf d}_m]^*$.
The sequence  $\gamma$ corresponds to the word $\langle \gamma\rangle=
\langle {\bf b}_0\rangle\langle {\bf b}_1\rangle\cdots \langle {\bf b}_l\rangle$.
In particular, we use $\gamma[j]$, projection of $\gamma$ on the $j$-th coordinate,
 to denote the word ${\bf b}_0[j]{\bf b}_1[j]\cdots {\bf b}_l[j]$.

Now, we define the set
$\Gamma$ of all the sequences $\gamma\in [{\bf d}_0][{\bf d}_1]^*[{\bf d}_m]^*$ satisfying
$\gamma[j]\in r_j$,  for each $1\le j\le k$, and $\gamma[1]\cdots \gamma[k]\in L'$.
We write $\langle \Gamma\rangle=\{\langle \gamma\rangle: \gamma\in\Gamma\}$.
We can claim:
(a). $\Gamma$ is a regular  language (on alphabet consisting of tuples
in $\cup_i [{\bf d}_i]$), and
(b).
$\langle \Gamma\rangle$ is also a regular language. The claims are not hard to show
(whose proof is omitted.).

However, the correspondence  between $\gamma$ and
$\langle \gamma\rangle$
is many-to-one (for the given and constant ${\bf d}_0,\cdots,{\bf d}_m$ defined in the ${\bf D}$).
More specifically,  each $\langle \gamma\rangle$ may correspond to $O(n^m)$ (where $m$ is the constant defined in ${\bf D}$, and
$n$ is the length of word $\langle \gamma\rangle$)
many $\gamma'$ with $\langle \gamma\rangle=\langle \gamma'\rangle$.
However, we  shall also notice that all the $\langle \gamma\rangle$'s form exactly $L$
In other words, each word $w\in L$  with length $n$ correspond to at most
$O(n^m)$ many distinct $\gamma$'s with $\langle \gamma\rangle=w$.

We now make a further translation on each word $\gamma$, in the form of,
${\bf b}_0{\bf b}_1\cdots {\bf b}_l$,
in $\Gamma$, into a new word $\hat\gamma$ in the form of
$({\bf b}_0, {\bf b}^0)({\bf b}_1, {\bf b}^1) \cdots ({\bf b}_l, {\bf b}^l)$,
where each newly added ${\bf b}^i$ is a $k$-arity tuple of symbols in $\Sigma^k$.
We further require that the translated $\hat\gamma$  is consistent in the following sense:
\begin{itemize}
\item Suppose that ${\bf b}^0=(b_1,\cdots, b_k)$. Then, each $b_j$ is the last symbol of the non-null word
${\bf b}_0[j]$.  (Recall that we assume that the vector  ${\bf d}_0$  is positive.)
\item  For each $i>0$, the following holds.  Consider $({\bf b}_{i-1}, {\bf b}^{i-1})$ and
 $({\bf b}_{i}, {\bf b}^{i})$. For each $j$,
\begin{itemize}
 \item  if  ${\bf b}_{i}[j]$ is non-null, then ${\bf b}^{i}[j]$ is the last symbol of the word ${\bf b}_{i}[j]$;
\item  if  ${\bf b}_{i}[j]$ is the null word, then ${\bf b}^{i}[j]={\bf b}^{i-1}[j]$.
\end{itemize}
\end{itemize}
So, the purpose of these newly introduced symbols in ${\bf b}^i$ are to ``memorize''
the symbols  that are most recently ``read'' on the $k$ ``tracks''.
After each word $\gamma$ in $\Gamma$ is translated into $\hat \gamma$, we obtain $\hat\Gamma$.
Clearly,  $\hat\Gamma$ is also regular (on alphabet $(\cup_i [{\bf d}_i])\times \Sigma^k$) and the translation
between $\gamma\in\Gamma$ and $\hat\gamma\in\hat\Gamma$ is
one-to-one and length-preserving.

We now do a final translation from
$\hat\Gamma$ into a new  language $\dot\Gamma$
such that there is a one-to-one and length-preserving correspondence between
$\dot\Gamma$ and $\langle \Gamma\rangle$. Consider
a word $\hat\gamma\in\hat\Gamma$ in the form of
$({\bf b}_0, {\bf b}^0)({\bf b}_1, {\bf b}^1) \cdots ({\bf b}_l, {\bf b}^l)$.
We shall emphasize again that each
 ``symbol'' in the word  is drawn from the product set $(\cup_i [{\bf d}_i])\times \Sigma^k$
and hence $\hat \gamma$ has length $l$.
We now stretch every such symbol longer by
padding it with some new ``stutter'' symbol ``${\heartsuit}$''
(actually we introduce many such symbols,  shown below).
For each symbol $({\bf b}_j, {\bf b}^j)$ in $\hat\gamma$,
 we translate into a word
$({\bf b}_j, {\bf b}^j)\heartsuit_{({\bf b}_j, {\bf b}^j)}^{|{\bf b}_j|-1}$, with length
$|{\bf b}_j|$,  where
$|{\bf b}_j|$ is the ``size'' of ${\bf b}_j$, which is the sum of the lengths of all elements (each element is a word by definition)
 in the vector ${\bf b}_j$.
Notice that the stutter symbol $\heartsuit_{({\bf b}_j, {\bf b}^j)}$ ``memorizes'' the vector ${\bf b}_j$, as well
as the most recently ``read'' symbols in ${\bf b}^j$.
(The definition is valid since $|{\bf b}_j|-1\ge 0$ as the vectors  ${\bf d}_i$ are all non-zero as we have mentioned earlier.)
After each symbol in  $\hat\gamma$ is translated, we obtain
the $\dot\gamma$.
After each word $\hat\gamma\in\hat\Gamma$ is translated into $\dot\gamma$, we obtain the language
$\dot\Gamma$.
It is straightforward to check that
$\dot\Gamma$ is regular (since $\hat\Gamma$ is regular) and
the translation from words $\langle \gamma\rangle$
all the way to $\dot\gamma$ is
 length-preserving and one-to-many:
\begin{quote} (*)
each $w\in L$  with length $n$ correspond to at most
$O(n^m)$ many distinct $\dot\gamma$'s with $\langle \gamma\rangle=w$.
\end{quote}

We now assign a cost function ${\bf U}$ to $\dot\Gamma$:
\begin{itemize}
\item  ${\bf U}(\heartsuit_{{\bf b}_1, {\bf b}^1}, \heartsuit_{{\bf b}_1, {\bf b}^1})=0$ for each ${\bf b}_1\in \cup_i [{\bf d}_i]$ and
${\bf b}^1\in \Sigma^k$;
\item ${\bf U}(({\bf b}_1, {\bf b}^1), \heartsuit_{{\bf b}_1, {\bf b}^1})=K_1+\cdots+K_k$, for each ${\bf b}_1\in \cup_i [{\bf d}_i]$ and
${\bf b}^1\in \Sigma^k$, where
 each $K_j$ is defined as follows.
$K_j=0$ if ${\bf b}_1[j]$ is the null word.
Otherwise, $K_j=(U)({\bf b}_1[j])$, where $U$ is the cost function given on $L$ in the theorem;
(Note that $(U)(a)=0$ by definition when $a$ is a single symbol in $\Sigma$.)
\item ${\bf U}(\heartsuit_{{\bf b}_1, {\bf b}^1}, ({\bf b}_1, {\bf b}^1))=K_1+\cdots+K_k$, for each ${\bf b}_1\in \cup_i [{\bf d}_i]$ and
${\bf b}^1\in \Sigma^k$, where
 each $K_j$ is defined as follows.
$K_j=0$ if ${\bf b}_1[j]$ is the null word.
Otherwise, $K_j=U({\bf b}^1[j], {\bf b}_1[j][1])$, where
${\bf b}_1[j][1]$ is the first symbol in the word ${\bf b}_1[j]$;
\item ${\bf U}(({\bf b}_1, {\bf b}^1), ({{\bf b}_2, {\bf b}^2}))=K_1+\cdots+K_k$, for each ${\bf b}_1, {\bf b}_2\in \cup_i [{\bf d}_i]$ and
${\bf b}^1, {\bf b}^2\in \Sigma^k$, where each $K_j$ is defined as follows.
\begin{itemize}
\item if ${\bf b}_2[j]$ is the null word, then $K_j=0$;
\item if otherwise, then $K_j=U({\bf b}^1[j], {\bf b}_2[j][1])$;
\end{itemize}
\end{itemize}
The construction of the cost function ${\bf U}$ is carefully designed so that the following property is guaranteed:
for every $w\in L$ that is long enough (more precisely,
 longer than the sum, which is a constant,  of all elements in the positive vector ${\bf d}_0$ in the definition of ${\bf D}$)
and every $\dot\gamma\in\dot\Gamma$ with $\langle \gamma\rangle=w$, the total cost $(U)(w)$ on $w$ equals
the total cost $({\bf U})(\dot\gamma)$ on $\dot\gamma$.
The ``long enough'' condition guarantees that the length of $\dot\gamma$ is at least 2.
By default, $({\bf U})(\dot\gamma)$ is defined to be 0 when the $\dot\gamma$ is a single symbol.
Recall that, even though the lengths of $w$ and $\dot\gamma$ are the same, the correspondence from
$w$ to $\dot\gamma$  is one-to-many (see the statement (*) mentioned earlier).
To sum up,  we finally have,
\begin{equation}
\limsup_{n\to \infty} {1\over n} \ln \sum_{w\in L, |w|=n}
 O(n^m)\cdot  e^{(U)(w)}  \ge \limsup_{n\to \infty} {1\over n} \ln \sum_{\dot\gamma\in \dot \Gamma, |\dot\gamma|=n}
   e^{({\bf U})(\dot\gamma)},
\end{equation}
and (since $\dot\gamma$'s in $\dot\Gamma$ with length $n$ are more than $w$'s in $L$ with length $n$)
\begin{equation}
\limsup_{n\to \infty} {1\over n} \ln \sum_{w\in L, |w|=n}
   e^{(U)(w)}  \le \limsup_{n\to \infty} {1\over n} \ln \sum_{\dot\gamma\in \dot \Gamma, |\dot\gamma|=n}
   e^{({\bf U})(\dot\gamma)}.
\end{equation}
Hence, we immediately have ${\rm G}_U(L)={\rm G}_{\bf U}(\dot\Gamma)$, noticing that
$\lim_{n\to \infty} {1\over n} \ln O(n^m)=0$ for the constant $m$.
The result then follows since, as we have mentioned earlier, $\dot\Gamma$ is a regular language and
its  free energy  ${\rm G}_{\bf U}(\dot\Gamma)$ is then computable (Theorem \ref{eqq0212}(2) and then Theorem \ref{EC}).
\end{proof}

\textbf{Proof of Theorem \ref{theorem11} }

\begin{proof}
We only show the result when $L=L'\cap L_{P,{\bf r},U}$ and
$P$ takes the form
$k_1\theta_1=k_2\theta_2=\cdots=k_k\theta_k$
where each $k_i$ is a positive integer constant; all other forms of the $P$ can be shown by generalizing the ideas in the following proof.

We first translate a tuple $\gamma=(w_1,\cdots,w_k)$ such that each $w_i\in r_i$,
\begin{equation}\label{equal}
k_1 (U)(w_1)=\cdots=k_k (U)(w_k),
\end{equation}
and $w_1\cdots w_k\in L'$  (we use $\Gamma$ to denote the set of all such $\gamma$'s)
 into
a word $\hat\gamma=(\hat w_1,\cdots,\hat w_k)$ as follows.
WLOG, we assume that each $w_i$ has length at least 2.
Each $\hat w_i$ is ``stretched'' from $w_i$ in the following sense. Suppose that
$w_i=b_0\cdots b_m$ for some $m\ge 2$.
Then, for each $b_j$, $1\le j<m$,
we replace $b_j$ with $b_j\heartsuit^{k_i\cdot U(b_j,b_{j+1})-1}$
(which has length $k_i\cdot U(b_j,b_{j+1})$). Hence, after the translation,
All the $\hat w_i$'s  share the same length $k_i\cdot (U)(w_i)+1$
(see (\ref{equal})). That is, we use the length of each $\hat w_i$ to ``memorize'' the total cost, multiplied by $k_i$, on $w_i$.

Then we translate each $\hat\gamma$ into
a word $\dot\gamma$
where the $j$-th symbol $\dot\gamma[j]$
of the word is the tuple $(w_1[j],\cdots,w_k[j])$ of the $j$-th symbols for all $w_i$'s.
$\dot\gamma$  will be perfectly aligned; i.e., each symbol is in $(\Sigma\cup\{\heartsuit\})^k$.
Here comes a  key step of this proof.
We drop all symbols in the form of $(\heartsuit,\cdots,\heartsuit)$ from $\dot\gamma$.
The result is denoted by $\bar\gamma$.   One shall observe that there are
 no two distinct $\gamma_1$ and $\gamma_2$ in $\Gamma$ such that
$\bar\gamma_1=\bar\gamma_2$. Hence, the translation from $\gamma$ to $\bar\gamma$ is one-to-one.

We now make a final translation.  For each symbol
$(b_1,\cdots,b_k)$ in $\bar\gamma$, by definition, no all $b_i$'s are $\heartsuit$.
For a symbol $b\in \Sigma\cup\{\heartsuit\}$, if it is in $\Sigma$, we define
$[b]_i$ to be the symbol $(\heartsuit,\cdot,\heartsuit, b, \heartsuit,\cdot,\heartsuit)\in (\Sigma\cup\{\heartsuit\})^k$
where the $b$ appears at the $i$-th position.
If, however, the $b$ is $\heartsuit$, we simply define $[b]_i$ to be the null word.
Now, each symbol $(b_1,\cdots,b_k)$ in $\bar\gamma$ is translated into
a non-null word $[b_1]_1\cdots [b_k]_k$. The resulting $\bar\gamma$ is denoted by $\tilde \gamma$.
Observe that the length of $\tilde \gamma$ is exactly the ``length''  $|w_1|+\cdots+|w_k|$ of $\gamma$.
We use $\tilde \Gamma$ to denote all the translated $\tilde\gamma$ from all the $\gamma\in\Gamma$.
It is not hard to show that: (a).  $\tilde \Gamma$ is a regular language, and (b).  the translation from
$\gamma\in\Gamma$ all the way to $\tilde\gamma\in \tilde \Gamma$ is one-to-one and length-preserving.

We can reuse the idea in the proof of Theorem \ref{linearl}
 of renaming $\heartsuit$ such that the most recently ``read'' non-$\heartsuit$ symbol in the same row is memorized
in the renamed symbol.
More precisely, suppose that  $\tilde\gamma$ is ${\bf b}_1\cdots{\bf b}_m$ for some $m$ and
each ${\bf b}_j\in  (\Sigma\cup\{\heartsuit\})^k$. The
 $i$-th row of $\tilde\gamma$ is then the word ${\bf b}_1[i]\cdots{\bf b}_m[i]$
where each ${\bf b}_j[i]$ is the $i$-th element in the tuple ${\bf b}_j$.
By definition, the $i$-th row contains at least two non-$\heartsuit$ symbols.
We now rename all the $\heartsuit$'s in the row as follows.
For each $\heartsuit$ that is after a symbol $b\in\Sigma$ (but there is no other symbols in $\Sigma$ in between)
we rename it into $\heartsuit_b$.  We leave all the $\heartsuit$'s at the beginning of the row that is not after any
symbol in $\Sigma$ unchanged.
After we rename each row in $\tilde\gamma$, we use $\ddot\gamma$ to denote the result
and use $\ddot\Gamma$ to denote all $\ddot\gamma$'s translated from all $\gamma\in\Gamma$.  Notice that
the translation from $\tilde\gamma$ to $\ddot\gamma$ is one-to-one and length-preserving.

Now, we are ready to show the result by first defining a cost function $u$ that generalizes the given $U$:
\begin{itemize}
\item  $u(\heartsuit_{b_1},b_2)=U(b_1, b_2)$ where $b_1,b_2\in\Sigma$;
\item  $u(\heartsuit_{b_1},\heartsuit_{b_1})=0$ where $b_1\in\Sigma$;
\item    $u(b_1,\heartsuit_{b_2})=0$ where $b_1,b_2\in\Sigma$;
\item    $u(\heartsuit, {b_2})=0$ where $b_2\in\Sigma$;
\item    $u(\heartsuit,\heartsuit)=0$.
\end{itemize}
(To make $u$ total, for all other cases it takes value 0.)
Now for any two ${\bf b}_1, {\bf b}_2\in  (\Sigma\cup\{\heartsuit\})^k$, we define
${\bf U}({\bf b}_1, {\bf b}_2)=u({\bf b}_1[1],{\bf b}_2[1])+\cdots+u({\bf b}_1[k],{\bf b}_2[k])$.
Clearly,  for each $\gamma=(w_1,\cdots,w_k)\in\Gamma$,
$(U)(w_1)+\cdots+(U)(w_k)=({\bf U})(\ddot\gamma)$. Hence, the total cost is also preserved by ${\bf U}$.
Notice that each $w=w_1\cdots w_k$ can only correspond to at most $O(n^k)$ may distinct $\gamma$'s, where $n=|w|$.
Using the similar argument earlier in the proof of Theorem \ref{linearl},
we have (notice that the absolute difference between $(U)(w)$ and
$(U)(w_1)+\cdots+(U)(w_k)$ is bounded by a constant):
\begin{equation}
\limsup_{n\to \infty} {1\over n} \ln \sum_{w\in L, |w|=n}
 O(n^k)\cdot  e^{(U)(w)}  \ge \limsup_{n\to \infty} {1\over n} \ln \sum_{\ddot\gamma\in  \ddot\Gamma, |\ddot\gamma|=n}
   e^{({\bf U})(\ddot\gamma)},
\end{equation}
and (since $\ddot\gamma$'s in $\ddot\Gamma$ with length $n$ are more than $w$'s in $L$ with length $n$)
\begin{equation}
\limsup_{n\to \infty} {1\over n} \ln \sum_{w\in L, |w|=n}
   e^{(U)(w)}  \le \limsup_{n\to \infty} {1\over n} \ln \sum_{\ddot\gamma\in \ddot \Gamma, |\ddot\gamma|=n}
   e^{({\bf U})(\ddot\gamma)}.
\end{equation}
Hence, we immediately have ${\rm G}_U(L)={\rm G}_{\bf U}(\ddot\Gamma)$, noticing that
$\lim_{n\to \infty} {1\over n} \ln O(n^k)=0$ for the constant $k$.
The result then follows since $\ddot\Gamma$ is clearly a regular language and
its pree energy  ${\rm G}_{\bf U}(\ddot\Gamma)$ is then computable (Theorem \ref{eqq0212}(2) and then Theorem \ref{EC}).
\end{proof}

\textbf{Proof of Theorem \ref{thmupper}}

\begin{proof}
WLOG, we assume that $M$ is strongly connected. In particular, we are only interested in
measuring the growth rate of the numbers of  initialized runs (as well as the input words on the runs), we can simply make
every state in $M$ an accepting state while doing this without affecting the numbers. This is because the runs and the input words
aforementioned are prefix closed.

For the given NFA $M$, let $V$ be the aforementioned cost function assigned on transitions $t$ in $M$.
For ease of presentation, we use ${\bf V}(t)$ to denote $e^{V(t)}$.
We first define notations.
Let $\alpha$ be any  word of length $n$.
We write $\alpha\leftarrow\tau$ if $\tau$ is a run of $M$ on $\alpha$ (note that $\tau$ does not  necessarily start from the initial state).
We write $p\prec \tau$ if $\tau$ starts from state $p$. As usual, $|\tau|$ is the number of transitions in $\tau$.
We simply write $(p,a,q)\in M$ if $(p,a,q)$ is a transition in $M$.
In particular, by ${\bf V}(\tau)$, we mean
${\bf V}(t_1)\cdot {\bf V}(t_2)\cdots {\bf V}(t_m)$, where $\tau=t_1,\cdots,t_m$ for some $m$.

We first claim, for each state $p$, and each  $\alpha$ with length $n$,
\begin{equation}\label{new9012}
\sum_{\substack{\tau: |\tau|=n, \\ \alpha\leftarrow\tau,p\prec \tau}} ({\bf V})(\tau)   ~\ge~
(  \sum_{\substack{\tau: |\tau|=n, \\ \alpha\leftarrow\tau,p\prec \tau}}  1  )^2
\end{equation}
We prove the claim by induction on $n$.

Base case ($n=1$).  In this case,   (\ref{new9012}) becomes, for each $p$ and $\alpha=a\in\Sigma$,
\begin{equation}\label{new9013}
\sum_{\substack{p': (p,a,p')\in M }}  k(p,a)  ~\ge~
(  \sum_{\substack{ p': (p,a,p')\in M}}  1  )^2,
\end{equation}
which holds since both LHS and RHS are equal to $k^2(p,a)$, by the definition of $k(p,a)$.

Induction.   We assume that (\ref{new9012})  holds for $n$.  Then, consider a word $a\alpha$, for any fixed $a\in\Sigma$, where
$\alpha$ is of length $n$. Now, the LHS of (\ref{new9012}) becomes
$$\sum_{\substack{\tau: |\tau|=n+1,  \\  a\alpha\leftarrow\tau,p\prec \tau}} ({\bf V})(\tau),$$
 which is, by definition,
$$\sum_{\substack{p': (p,a,p')\in M }}  k(p,a) \cdot \sum_{\substack{\tau: |\tau|=n,  \\  \alpha\leftarrow\tau,p'\prec \tau}} ({\bf V})(\tau) .$$
Using induction hypothesis, we therefore have,
$$\sum_{\substack{\tau: |\tau|=n+1,\\  a\alpha\leftarrow\tau,p\prec \tau}} ({\bf V})(\tau) ~\ge ~
\sum_{\substack{p': (p,a,p')\in M }}  k(p,a) \cdot (\sum_{\substack{\tau: |\tau|=n,\\  \alpha\leftarrow\tau,p'\prec \tau}} 1)^2
$$
$$\ge~ k(p,a)\cdot \sum_{\substack{p': (p,a,p')\in M }}   (\sum_{\substack{\tau: |\tau|=n, \\ \alpha\leftarrow\tau,p'\prec \tau}} 1)^2$$
(using Jensen's inequality $\sum_N z_i^2\ge {1\over N}\cdot {{(\sum_N z_i)^2}}$)
$$\ge~ k(p,a)\cdot {1\over {k(p,a)}}\cdot
 {  { ( \sum_{\substack{p': (p,a,p')\in M }} \sum_{\substack{\tau: |\tau|=n, \\ \alpha\leftarrow\tau,p'\prec \tau}} 1  )^2 }                         }$$
(noticing that, in above, $k(p,a)$ is the number of $p'$ satisfying $(p,a,p')\in M$)
$$=~( \sum_{\substack{\tau: |\tau|=n+1,\\  a\alpha\leftarrow\tau,p\prec \tau}}  1  )^2.$$
Hence, the claim follows.

Let $p_0$ be the initial state of $M$.
Now, we establish for  each $n$,
\begin{equation}\label{new9014}
( \sum_{\substack{\tau: |\tau|=n,\\ p_0\prec \tau}} {\bf V}(\tau) )   \cdot ( \sum_{\substack{\alpha: ~\exists \tau. ~ |\tau|=n, \\ \alpha\leftarrow\tau,p_0\prec \tau}} 1    )  ~\ge~ (  \sum_{\substack{\tau: |\tau|=n,\\  p_0\prec \tau}}  1 )^2.
\end{equation}
Let $K$ be the set of distinct $\alpha$ satisfying $\exists \tau.  ~|\tau|=n, \alpha\leftarrow\tau,p_0\prec \tau$.  The size of the set is denoted by $|K|$.
Now, the LHS of (\ref{new9014}) is written
$$ ( \sum_{\alpha\in K}\sum_{\substack{\tau:  |\tau|=n, \\ \alpha\leftarrow\tau,p_0\prec \tau        }         }    {\bf V}(\tau)           )\cdot |K     |,         $$
which is, from the previous claim in (\ref{new9012}),
$$ \ge~ \sum_{\alpha\in K}    (  \sum_{\substack{\tau:  |\tau|=n, \\ \alpha\leftarrow\tau,p_0\prec \tau        }         }   1 )^2\cdot   |K     |        $$
(using the Jensen's inequality again)
$${1\over {   |K     | }}   (   \sum_{\alpha\in K} \sum_{\substack{\tau:  |\tau|=n, \\ \alpha\leftarrow\tau,p_0\prec \tau        }         }   1     )^2\cdot   |K     |                $$
$$= (    \sum_{\substack{\tau:  |\tau|=n, \\ p_0\prec \tau        }         }   1            )^2,  $$
which is (\ref{new9014}).

Now, from (\ref{new9014}),
it is direct to have,
$$ {1\over n}\ln    \sum_{\substack{\tau:  |\tau|= n, \\ p_0\prec \tau        }         }  {\bf V}(\tau)      ~-~   {1\over n}\ln    \sum_{\substack{\tau:  |\tau|= n, \\ p_0\prec \tau        }         }    1     $$
$$\ge~  {1\over n}\ln    \sum_{\substack{\tau:  |\tau|=n, \\ p_0\prec \tau        }         }    1  ~-~ {1\over n}\ln    \sum_{\substack{\alpha: ~\exists \tau.~ |\tau|=n, \\ \alpha\leftarrow\tau, p_0\prec \tau        }         }    1 .   $$
Taking limit on $n$ (all the four limits exist -- using the assumption mentioned at the beginning of the proof --  and using
(\ref{eqq8975})), we immediately have,
$\lambda^+_M={\cal E}(M_V)-{\cal E}(M_{\bf 0})\ge \lambda_M$.
The result follows.
\end{proof}




\end{document}